\documentclass[%
reprint, prd
]{revtex4-2}

\usepackage{graphicx}
\usepackage{dcolumn}
\usepackage{bm}
\usepackage{tikz}

\begin{document}

\preprint{APS/123-QED}

\title{Path Sums for Propagators in Causal Sets}

\author{Samuel Shuman}%
\email{shumans@oregonstate.edu}
\affiliation{Department of Physics, Oregon State University.}

\date{\today}

\begin{abstract}
A major challenge in Causal Set research is that theories need only to match general relativity and quantum field theory in the appropriate limits. This means that there should be many different ways to calculate a scalar field propagator in a causal set that match the known limits, but may give significantly different results on the small scale. In this work, we explore under what conditions a path sum will correspond to a scalar field propagator in such a way that it matches the known value in the continuum limit. A family of solutions for the path sum is found and is verified numerically in a few specific cases.
\end{abstract}

\maketitle


\section{\label{sec:intro} Introduction}

\subsection{\label{sec:background1}Background: Causal Sets}

Causal Set Theory (CST) is a candidate theory of quantum gravity. Like many such theories, it focuses on describing the Planck scale structure of spacetime in a way that is mostly consistent with general relativity and quantum field theory. As one might expect, there are many ways to approximate these theories on the Planck scale, so to make progress, we must decide which classical properties should be fundamental aspects of the new theory and which should be emergent on the large scale. CST is based on the idea that spacetime is fundamentally discrete, and has a fundamental causal structure, but Lorentzian geometry is emergent.

The central idea of CST, that causal structure and discreteness are enough to recover Lorentzian geometry on the large scale, has its foundation in a series of papers in the 1970s. These papers showed that for all past and future distinguishing spacetimes the causal structure is enough to determine the conformal geometry~\cite{1,2}. This means that if you know the causal structure of a spacetime and the volume of every region, that is enough to recover the entire geometry. As summarized by Rafael Sorkin in~\cite{16}:

\[\text{Causal Structure} + \text{Volume} = \text{Geometry}\]
As we will see, discreteness, when treated in the ways we will describe, can tell you about the spacetime volume of any region, with the number of points in a region corresponding to the volume. This correspondence depends on a parameter called the density, $\rho$, of the causal set. Thus we get:

\[\text{Causality} + \text{Discreteness} \approx \text{Geometry}\]
Therefore, a discrete set of events with a causal ordering should be enough to recover the geometry of a Lorentzian manifold. This idea was first laid out by Bombelli, Lee, Meyer, and Sorkin in 1987~\cite{7}.

\subsubsection{Definitions}
What follows are some important definitions in CST. A \textbf{Causal Set} is a set of events, $C$, paired with a parameter, $\rho$, called the density, and an ordering relation (the causal order), $\prec$, satisfying the following properties:
\begin{itemize}
    \item Transitivity $x \prec y$, $y \prec z$ $\Rightarrow x \prec z$
    \item Anti-symmetry $x \prec y$ $\Rightarrow y \not\prec x$
    \item Local-finiteness
\end{itemize}
Local-finiteness requires that the \textbf{interval}
\[[x,y] = \{z \| x \prec z \prec y\}\]
is finite for all $x,y \in C$. Note that in the context of general relativity, the interval is often referred to as a causal diamond. To summarize what these requirements correspond to physically, transitivity tells us that this relation orders the causal set and can be interpreted as a casual structure, antisymmetry tells us that there are no closed causal loops $x \prec y \prec x$, and local-finiteness tells us that regions of a causal set that correspond to finite regions in a manifold description should have finitely many elements. This last requirement guarantees that our causal set is discrete and that there can exist a number-volume correspondence as mentioned above.

We must also consider a few types of \textbf{trajectories} in causal sets. A trajectory is a sequence of events in $C$. Note that no requirement is made here regarding causal structure. A \textbf{chain} is a sequence of events $(x_n)$ in $C$ such that $x_i \prec x_{i+1}$ for all $i$. We say $x$ is \textbf{linked} to $y$, denoted $x \prec * y$, if $x \prec y$ and the interval $[x,y]$ is empty. We can then define a \textbf{path} as a sequence $(x_n)$ in $C$ such that $x_i \prec * x_{i+1}$ for all $i$. To state this a different way, paths are chains that are as close to continuous as possible. The length of a trajectory $\{x_0, \dots, x_n\}$ is defined to be $n$. Finally, we will define a \textbf{jump} between two events in a spacetime to be a trajectory of length 1. In particular, this means the jump from $x$ to $y$ for some $x, y \in C$ is the sequence $\{x,y\}$ with no intermediate points.

\subsubsection{Sprinklings}
The dynamics of CST have not been fully determined, so we cannot yet solve for the causal order corresponding to physical conditions from first principles. In this paper, we will need to generate causal sets that correspond to a flat spacetime. To do this we will use a method called sprinkling. Sprinkling is a strategy to generate a causal set corresponding to a given manifold by taking a random set of points in that manifold as our events and using the causal structure of the manifold to define the causal relation.

\begin{figure}[ht]
    \centering
    \includegraphics[width=8.6cm]{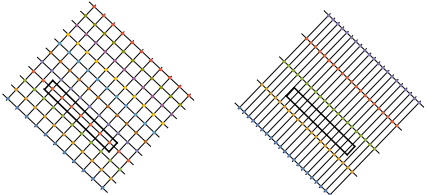}
    \caption{From ~\cite{3}, a lattice spacetime in two dimensions. The number-volume correspondence only holds in a specific frame and fails to hold in a Lorentz boosted frame.}
    \label{fig:figure1}
\end{figure}

One might expect a regular lattice to be a more appropriate sprinkling procedure than randomly selecting events. However, a random sprinkling is necessary to preserve Lorentz symmetry. To make sense of this, note that a regular lattice in a spacetime defines a preferred reference frame (see figure \ref{fig:figure1}). We will see that a Poisson process is a natural choice for this random distribution.

A Poisson process in a 4-dimensional spacetime is defined analogously to a 1-dimensional Poisson distribution. It is defined by a single parameter that tells you the density at which events are selected. In general, a Poisson process must satisfy two properties. One is that the probability of finding exactly $n$ points in a region of volume $V$ is:
\begin{equation}
P\{N = n\} = \frac{(\rho V)^n}{n!}e^{-\rho V}
\end{equation}
Here, $\rho$ is the average number density of points in the spacetime. The other required property is that the number of points in disjoint, bounded regions is independent.

\begin{figure}[ht]
    \centering
    \includegraphics[width=8.6cm]{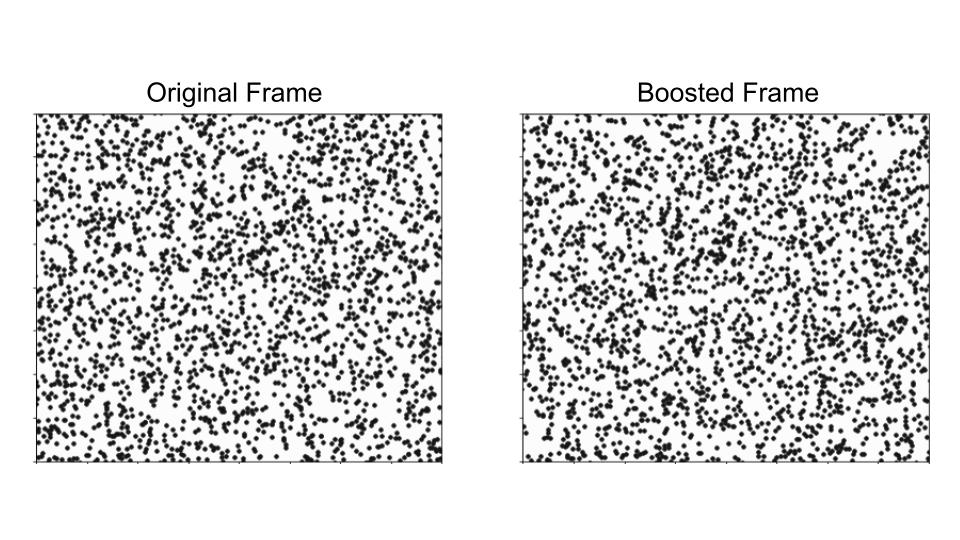}
    \caption{A Poisson sprinkling in a flat 2-dimensional spacetime shown in both the original reference frame and one that is Lorentz boosted. This was calculated with $\rho = 600$ and a relative velocity of $v = 0.6c$. Unlike the lattice sprinkling shown in figure 1, the Poisson random sprinkling has no preferred reference frame.}
    \label{fig:figure2}
\end{figure}

This is what is called a homogeneous Poisson process, meaning that the rate at which events occur, $\rho$, is constant and does not depend on the location in spacetime. Homogeneous Poisson processes are particularly easy to simulate in finite regions. First, use the Poisson distribution and the volume of the region to randomly determine the number of points that will be sprinkled. Then each point is placed in the region with a uniform random distribution. For rectangular regions, this can be done by selecting each coordinate from a one dimensional uniform distribution~\cite{21}.

\subsubsection{Correspondences}
If CST is correct, then the underlying structure of the universe is a causal set and any manifold representation of spacetime is just an approximation. That being said, the manifold approximation for spacetime must be very good when we ``zoom out" from the discreteness scale.

This leads to the ``Fundamental Conjecture of CST" or the ``Hauptvermutung"~\cite{3}. In short, if a causal set that is generated by a Poisson sprinkling into a manifold $M$ can also be achieved as a Poisson sprinkling into a manifold $M'$ at the same density, then the two manifolds $M$ and $M'$ must have almost identical geometries. There have been some suggestions, such as in~\cite{5}, to make this statement more rigorous but there is not yet a clear consensus. If this conjecture was not true, then CST would need additional structure to recover general relativity on the large-scale.

One way this problem has been approached is to establish correspondences between properties of a causal set and properties of any manifold it can be faithfully embedded in. A manifold faithfully embeds a causal set if the causal set can be generated as a Poisson sprinkling on that manifold. The idea is to say that any manifold that faithfully embeds a particular causal set must approximate certain geometric properties on the large-scale. The most basic example is the number-volume correspondence we have already discussed. Since our points are sprinkled by a Poisson process, the fractional variance in the number of points in a region is:
\begin{equation}
    \frac{\delta n}{n} = \frac{\sqrt{n}}{n} = \frac{1}{\sqrt{n}}
\end{equation}
Thus, regions of the causal set containing a large number of events, $n$, can be known within a very small uncertainty to have volume $V = n/\rho$ in any manifold that faithfully embeds it.

Another correspondence is for the geodesic proper time separating two causally connected events. In 2 dimensions~\cite{13}, this proper time can be estimated as:
\begin{equation}
    \tau^6 = \frac{1}{8\rho^3}(J_1 - 2J_2 + J_3) \label{proper time}
\end{equation}
with $J_k$ defined by
\begin{equation}
    J_k = (2k+2)(2k+4)(2^3)\left(kC_k\right)^{3/k} \label{Js eqn}
\end{equation}
and $C_k$ defined as the number of chains of length $k$ connecting the events. These formulas vary slightly depending on the dimension of the spacetime.

It should be noted that many such correspondences have been investigated. There are correspondences to estimate the curvature, dimension, volume, proper time, etc of the regions of any manifold that can be represented by that causal set. See~\cite{3} for further discussion of these correspondences.

\subsection{\label{sec:background2}Background: Path Integrals and Propagators}
In general, a propagator describes how a quantum system transitions from one state to another. A common special case is a propagator that describes how a system transitions from one event to another. This will be proportional to the probability amplitude associated with transitioning between these events. For events $x,y$ in spacetime, we will denote this propagator $K(x,y)$. The functional form of the propagator depends on the quantum system being described.

The path integral formulation developed by Richard Feynman shows that some propagators between events in spacetime can be calculated by integrating over trajectories connecting the two events~\cite{22}. Different propagators can be found by including different types of trajectories in this integral, but not all propagators can be defined this way~\cite{10}. To find the total probability amplitude of transitioning between two events, we can assign a probability amplitude to each path and add up each of those contributions:
\begin{equation}
    K(x,y) = \int_x^y \delta q \hspace{4pt} e^{iS[q]} \label{path int}
\end{equation}
The integral must be taken over paths, $q$, connecting $x$ and $y$. The probability amplitude associated with each path is $e^{iS[q]}$ where $S[q]$ is the action of that path, which will depend on the quantum system being described.

\subsection{\label{sec:prev} Previous Work}
There has been past work considering scalar field propagators in causal sets. For example, in~\cite{8, 23} the author considered a model that assigned an amplitude to each jump along a path (called the hop amplitude $a$), and another to each intermediate event along the path (called the stop amplitude $b$). They were able to show that there are values of these constants that recover the free scalar field retarded propagator when averaged over sprinklings and considered in the appropriate continuum limit. The main idea is to create a matrix representation of the propagator that approximately matches the value of the propagator in the continuum.

Now let us consider an illustrative example for how the average values of such propagators were calculated. In~\cite{8, 23}, when summing over paths and averaging over sprinklings, the propagator becomes:
\begin{equation}
    K(x,y) = \sum_n a^nb^{n-1} P_n(x,y)    
\end{equation}
where $P_n(x,y)$ is the average number of paths of length $n$ from $x$ to $y$. This expression comes from the fact that every path of length $n$ has $n$ hops and $(n-1)$ stops. The author then set up an integral for $P_n(x,y)$ and used that to get an integral relationship for the propagator
\begin{eqnarray}
    P_n(x,y) = \rho^{n-1}\int dz_1 \int dz_2 \mathellipsis \int dz_{n-1} \nonumber \\
    \mu(x,z_1) \mu(z_1, z_2) \mathellipsis \mu(z_{n-1}, y)
\end{eqnarray}
Here $\mu(x,y)$ is the probability that events $x$ and $y$ are linked. This leads to the integral relation:
\begin{equation}
    K(y-x) = a \mu(y-x) + \rho ab\int dz \mu(z-x) K(y-z)
\end{equation}
Since the integral is a convolution, this equation can be solved with a Fourier transform.
\begin{equation}
    \tilde{K}(p) = \frac{a\tilde{\mu}(p)}{1 - \rho ab \tilde{\mu}(p)}
\end{equation}
Then one only has to Fourier transform back to get an expression for the average over sprinklings of the scalar field propagator associated with these hop and stop amplitudes.

While this work was able to suggest a method to construct a propagator on a causal set, there is an important consideration. Since the only restrictions are that the path sum matches the continuum propagator on average and does not vary too much over sprinklings, we should expect many different formulas for the propagator to match these conditions. For that reason, it is useful to solve this problem in a more general way, so that we might categorize a greater variety of possible path sums that are consistent with the continuum calculation.

A similar approach was taken in~\cite{9}, but in that paper the authors allowed for non-constant hop amplitudes and did not include stop amplitudes. While they solved for a matrix relationship between these hop amplitudes and the propagator in a way that is similar to what we will see in the next section, they did not discuss how this relationship would average over sprinklings.

\section{\label{sec:sums} Path Sums for Propagators}
In this section, we will develop a general method for analyzing the relationship between a scalar field propagator in a causal set and the jump amplitude matrix $T$ (which serves the same role as the matrix $A$ in~\cite{9}). By looking at this relationship after averaging over sprinklings into Minkowski space, we will require that we recover the same propagator as the continuum calculation. In~\cite{8}, there is discussion for how such a propagator could be used to define a scalar field theory on a causal set.

\subsection{\label{sec:jumps} Propagators and Jump Amplitudes \\ on Causal Sets}
First, since the causal set is discrete we can label events in the causal set by non-negative integer indices. Let us assume the causal set is finite. Then define the matrix $T$, by
\begin{eqnarray}
    T_{ij} \equiv \text{The probability amplitude associated with} \nonumber \\ 
    \text{jumping from event } i \text{ to event } j \nonumber
\end{eqnarray}
We will also define
\begin{eqnarray}
    \sigma_{ij} \equiv \text{The total probability amplitude of all} \nonumber \\
    \text{trajectories from event } i \text{ to event } j \nonumber
\end{eqnarray}
The propagator will be proportional to the matrix $\sigma$, but we must include a constant that incorporates the units of the propagator in the same way that $\delta q$ incorporates the units of the propagator in equation \ref{path int}. In particular, we will define $K_{xy} = a \sigma_{xy}$.

Since all trajectories can be broken down into a sequence of jumps, we should expect a relationship between $T$ and $\sigma$. Consider organizing trajectories from $x$ to $y$ by the first jump taken. Every trajectory from $x$ to $y$ is either a direct jump to $y$, or a jump to some element $z$ followed by a trajectory from $z$ to $y$. Therefore we have
\begin{equation}
    \sigma_{xy} = T_{xy} + \sum_z T_{xz} \sigma_{zy}
\end{equation}
Since $K_{xy} = a \sigma_{xy}$, we get the composition relation
\begin{equation}
    K_{xy} = a T_{xy} + \sum_z T_{xz}K_{zy} \label{comp law}
\end{equation}
Or, written in matrix form
\begin{equation}
    K = aT + TK
\end{equation}
\begin{equation}
    K = aT(I-T)^{-1}
\end{equation}
This derivation allows us to define the propagator whenever $(I - T)$ is invertible. If $(I - T)$ is not invertible, then there is not a clear way to define a propagator in terms of the jump amplitude matrix, $T$. Note that when the allowed jumps are all causal, $(I - T)$ is guaranteed to be invertible and so the propagator is well-defined ~\cite{8,23}. 

\begin{figure}[ht]
    \centering
    \begin{tikzpicture}
        \node [black] at (3,0){\textbullet};
        \node [left, black] at (3,0){$x$};
        \node [black] at (4.3,1.12){\textbullet};
        \node [right, black] at (4.3,1.12){$z$};
        \node [black] at (4,2){\textbullet};
        \node [black] at (4.5,2.96){\textbullet};
        \node [black] at (3,4){\textbullet};
        \node [left, black] at (3,4){$y$};
        \node [black] at (2.3,0.64){\textbullet};
        \node [black] at (1.5,1.6){\textbullet};
        \node [black] at (2.5,3.2){\textbullet};
        \draw [dashed, blue] (3,0) -- (3,4);
        \node [left, blue] at (3,2) {$T_{xy}$};
        \draw [dashed, blue] (3,0) -- (4.3,1.12);
        \node [blue] at (4,0.4) {$T_{xz}$};
        \draw [dashed, red] (4.3, 1.12) -- (4,2) -- (4.5,2.96) -- (3,4);
        \node [red] at (4.6,2.4) {$K_{zy}$};
    \end{tikzpicture}
    \caption{Every trajectory from $x$ to $y$ is either a direct jump to $y$, or a jump to some event $z$ followed by a trajectory from $z$ to $y$. This is reflected in the composition law $K = aT +TK$.}
    \label{fig:figure3}
\end{figure}
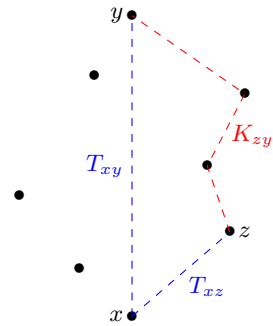
\subsection{\label{sec:average} Averaging Over Sprinklings}
The next step taken in previous attempts at deriving path sums in causal sets is equivalent to postulating a model $T$ matrix (such as the hop and stop amplitudes considered in~\cite{8,23}) and then carefully considering what integral relations may hold for this model. In this work, we will take a more general approach by defining an average value of the matrix $T$ and of the propagator.

First, let $x$ and $y$ be two elements in the background manifold $M$. Consider the space of all possible causal sets faithfully embedded in $M$ that include $x$ and $y$. If we label these two events with the indices 0 and 1 in these causal sets, then we will define
\begin{equation}
    T(x,y) \equiv \text{average over sprinklings of }T_{01} \nonumber
\end{equation}
\begin{equation}
    K(x,y) \equiv \text{average over sprinklings of }K_{01} \nonumber
\end{equation}
We will assume these averages lead to functions that are sufficiently smooth and continuous. This seems reasonable since the $K$ and $T$ matrices will be defined in terms of the causal structure of a sprinkling into a smooth manifold. Importantly, since we are doing this calculation in Minkowski spacetime, we will also assume translational invariance:
\[T(x,y) = T(y - x)\]
\[K(x,y) = K(y - x)\]
Next we will consider how these average functions follow the composition law in equation \ref{comp law}. For the $K_{01}$ matrix element we have
\begin{equation}
    K_{01} = aT_{01} + \sum_i T_{0i}K_{i1}
\end{equation}
When we average over sprinklings, $K_{01}$ becomes $K(y-x)$ and $T_{01}$ becomes $T(y-x)$. The sum will include all points in the causal set, so when averaged over all possible sprinklings we will need to include every point in the manifold. To make the next step more clear, consider the discreteness scale volume $V_0 = 1/\rho$, which can be thought of as the average volume associated with each event in a sprinkling. We can write the sum as
\begin{equation}
    \sum_i T_{0i}K_{i1} = \rho \sum_i T_{0i}K_{i1} V_0
\end{equation}
The sum now evaluates $T_{0i}K_{i1}$ at each event in the spacetime and multiplies by a volume associated with that event. Now we can see that this is a type of Riemann sum that will become an integral when we average over all sprinklings. Thus we get the average composition relation:
\begin{equation}
    K(y - x) = aT(y - x) + \rho \int dz T(z - x) K(y - z)
\end{equation}
Since the integral is a convolution, this relationship can be solved by Fourier transforming the equation.
\begin{equation}
    \tilde{K}(p) = a\tilde{T}(p) + \rho \tilde{T}(p)\tilde{K}(p)
\end{equation}
\begin{equation}
    \tilde{K}(p) = \frac{a \tilde{T}(p)}{1 - \rho \tilde{T}(p)}
\end{equation}
Then all that remains is to Fourier transform back to find the average propagator $K(y-x)$.

While this is similar to what was done in~\cite{8,23}, since we have not specified the jump amplitude function, we can solve for that instead.
\begin{equation}
    \tilde{T}(p) = \frac{\tilde{K}(p)}{a + \rho \tilde{K}(p)} \label{T Fourier}
\end{equation}
Then Fourier transforming back would tell us what jump amplitudes would be needed on average to recreate a given propagator as a path sum in a causal set.

\subsection{\label{sec:solving} Solving for Jump Amplitudes}
The Feynman, retarded, and advanced propagators for scalar fields all have Fourier transforms of the form
\begin{equation}
    \tilde{K}(p) = \frac{1}{f(p) + m^2}
\end{equation}
For example, for the Feynman propagator we would set $f(p) = p^2 - i\varepsilon$. Plugging this into equation \ref{T Fourier} yields:
\begin{equation}
    \tilde{T}(p) = \frac{1}{a}\cdot \frac{1}{f(p) + m^2 + \rho/a}
\end{equation}
This has the same form as the Fourier transform of the propagator but with the constant $(m^2 + \rho/a)$ taking the place of $m^2$. To simplify this expression, define the factor $\beta = \sqrt{1 + \frac{\rho}{m^2 a}}$. Then $m^2 \mapsto (m^2 + \rho/a)$ may instead be written as $m \mapsto (\pm \beta m)$. This gives a final result for the average value of the jump amplitudes for propagators of this form:
\begin{equation}
    T(y-x) = \frac{1}{a}K(y-x)\text{ , } m \mapsto \pm \beta m \label{T from K}
\end{equation}

\subsection{\label{sec:units} Units of $a$}
So far in this paper, we have used natural units with $\hbar = c = 1$. In this section we will reintroduce $\hbar$ in order to better understand how $a$ should depend on $\rho$ and $m$. The constant $a$ that appears in these expressions is included so that the propagator is not unitless. To determine what units we should expect for $a$, note that $a$ has the same units as the propagator, $K$. Consider the Feynman propagator. This has the Fourier transform:
\begin{equation}
    \tilde{K} = \frac{1}{p^2 + m^2}
\end{equation}
This is a Green's function for the Klein-Gordon equation, which is a statement of the relativistic energy relation $E^2 = p^2 + m^2$. Therefore we should expect $\tilde{K}$ to have the same units as $1/m^2$. Now consider the Fourier transform for $K$.
\begin{equation}
    \tilde{K} = \int dx e^{ipx} K
\end{equation}
Since $\tilde{K}$ has units of $1/m^2$ and $dx$ has units of spacetime volume, $K$ must have units of $\frac{1}{Vm^2}$. In order for these units to come directly from the constants $m$ and $\rho$, we must have a normalization of the form $a = \alpha \frac{\rho}{m^2}$, where $\alpha$ is a unitless parameter. This yields a $\beta$ factor of the form
\begin{equation}
    \beta = \sqrt{1 + \frac{1}{\alpha}}
\end{equation}

Since fundamental constants like $\hbar$ and $G$ may also contribute to the units of $a$, we cannot rule out that $\alpha$ may depend on $\rho$ and $m^2$. As we will discuss in section \ref{sec:num res}, the only clear restriction on $\alpha$ is that it is non-zero.

\section{\label{sec:apps} Applications}
In this section, we will apply the formulas derived in section \ref{sec:sums} to construct path sums for the retarded propagator and the Feynman propagator. First, we will show that this formulation is consistent with past work. Then we will use numerical simulations to show that these results are consistent with the continuum values for the retarded propagator. In this section it will be useful to define:
\begin{equation}
    \nu(y-x) \equiv \Theta(y^0 - x^0) \Theta\Big(\tau(y-x)^2\Big)
\end{equation}
Note that this is 1 if $x \prec y$ and 0 otherwise.

\subsection{\label{sec:results1} Results for the Retarded Propagator}

\subsubsection{Comparison to Previous Results}
In order to compare the following results, which are in terms of jump amplitudes, to the results in~\cite{8,23}, we must first consider how hop and stop amplitudes can be expressed in terms of jump amplitudes. All but the last jump along a trajectory consists of a hop and a stop (since only intermediate events count as stops). This means that if we multiply the hop and stop amplitudes from~\cite{8,23} we should get something that matches the jump amplitudes in the following calculations. Furthermore, since the last jump includes only a hop we must divide the trajectory's probability amplitude by the stop amplitude to get its contribution to the propagator. This means our unit constant $a$ should be the inverse of the stop amplitude. Note that the unit constant $a$ introduced in the previous section is not the same as the hop amplitude $a$ from~\cite{8,23}.

\subsubsection{Results in 2 Dimensions}
In 2 dimensions, the retarded propagator takes the form:
\begin{equation}
    K(y-x) = \nu(y-x)\frac{1}{2}J_0\Big(m\tau(y-x)\Big)
\end{equation}
Using equation \ref{T from K}, in order for the path sum to match this on average, the average jump amplitudes must be:
\begin{equation}
    T(y-x) = \frac{1}{a}\nu(y-x)\frac{1}{2}J_0\Big(\pm\beta m\tau(y-x)\Big) \label{T in 2D}
\end{equation}
The derivation discussed in~\cite{8} is equivalent to setting $a = -\frac{\rho}{m^2}$. When this choice is made we get:
\begin{equation}
    \beta = \sqrt{1 + \frac{\rho}{m^2a}} = 0
\end{equation}
\begin{equation}
    T(y-x) = -\frac{m^2}{\rho}\frac{1}{2}\nu(y-x)
\end{equation}
The $\Theta$-functions ensure that jumps are only allowed from $x$ to $y$ if $x \prec y$. This result then says a jump to such a future-connected event should have an amplitude of $-\frac{m^2}{2\rho}$. This is equivalent to the result of~\cite{8} where $-\frac{m^2}{2\rho}$ is the product of the hop and stop amplitudes.

\subsubsection{Results in 4 Dimensions}
In 4 Dimensions, the retarded propagator takes the form:
\begin{eqnarray}
    K(y-x) = \nu(y-x)\bigg(\frac{1}{2\pi}\delta\Big(\tau(y-x)^2\Big) \nonumber \\
    - \frac{m}{4\pi\tau(y-x)}J_1\Big(m\tau(y-x)\Big)\bigg)
\end{eqnarray}
This means the jump amplitudes must be of the form:
\begin{eqnarray}
    T(y-x) = \frac{1}{a}\nu(y-x)\bigg(\frac{1}{2\pi}\delta\Big(\tau(y-x)^2\Big) \nonumber \\
    - \frac{\pm \beta m}{4\pi\tau(y-x)}J_1\Big(\pm \beta m\tau(y-x)\Big)\bigg)
\end{eqnarray}
If we again set $a = -\frac{\rho}{m^2}$ and $\beta = 0$ we get:
\begin{equation}
    T(y-x) = -\frac{m^2}{\rho}\frac{1}{2\pi}\delta\Big(\tau(y-x)^2\Big)\nu(y-x)
\end{equation}
As discussed in~\cite{8}, we will need the result:
\begin{equation}
    \lim_{\rho\to\infty} \sqrt{\rho}\mu(y-x) = \frac{\sqrt{24}}{2}\delta\Big(\tau(y-x)^2\Big)\nu(y-x)
\end{equation}
Here I have again used $\mu(y-x)$ to represent the probability that $x$ is linked to $y$. Then in the high density limit we have:
\begin{equation}
    T(y-x) \approx -\frac{m^2}{\rho}\frac{\sqrt{\rho}}{\sqrt{24}\pi} \mu(y-x)
\end{equation}
This says that if we allow only jumps between linked events and give them the amplitude $-\frac{m^2}{\rho}\frac{\sqrt{\rho}}{\sqrt{24}\pi}$ then this will give the correct propagator on average. This is the same as the product of hop and stop amplitudes found in~\cite{8}.

\subsubsection{Numerical Results \label{sec:num res}}
While the previous calculations show that the causal set propagator should match the continuum propagator for any value of $a$ on average, it is conceivable that the variation may still be too large for the model to be acceptable. To test this, we will use numerical simulations.

To start, events in the causal set are chosen using a Poisson sprinkling with density $\rho = 4500$ into a causal diamond in a flat 2D spacetime. Then the causal order of the set is determined by assuming a Minkowski metric. The proper time separating each pair of events is estimated from the causal structure using equations \ref{proper time} and \ref{Js eqn}. These proper times are used to calculate a jump amplitude matrix:
\begin{equation}
    T_{xy} = C_{xy}\frac{m^2}{2\rho\alpha}J_0\left(\sqrt{1 + \frac{1}{\alpha}} m\tau_{xy}\right) \label{T Model}
\end{equation}
where $C_{xy} = 1$ if $x \prec y$ and $0$ otherwise. Note that this formula has the same form as the average jump amplitude from equation \ref{T in 2D}. In particular, this formula assumes $a$ is of the form $a = \alpha \frac{\rho}{m^2}$.

There is a slight issue with this formula for the jump amplitudes. Even though the average over sprinklings of $\tau_{xy}$ should be the manifold proper time $\tau(y-x)$, the average over sprinklings of equation \ref{T Model} will not always match equation \ref{T in 2D}. This is because the average of a function is not generally the same as the function of the average. However, we should expect this effect to go away as $\rho \to \infty$ since the variation over sprinklings in $\tau_{xy}$ goes to zero in the limit~\cite{13}.

\begin{figure*}[ht]
    \centering
    \includegraphics[width=17.2cm]{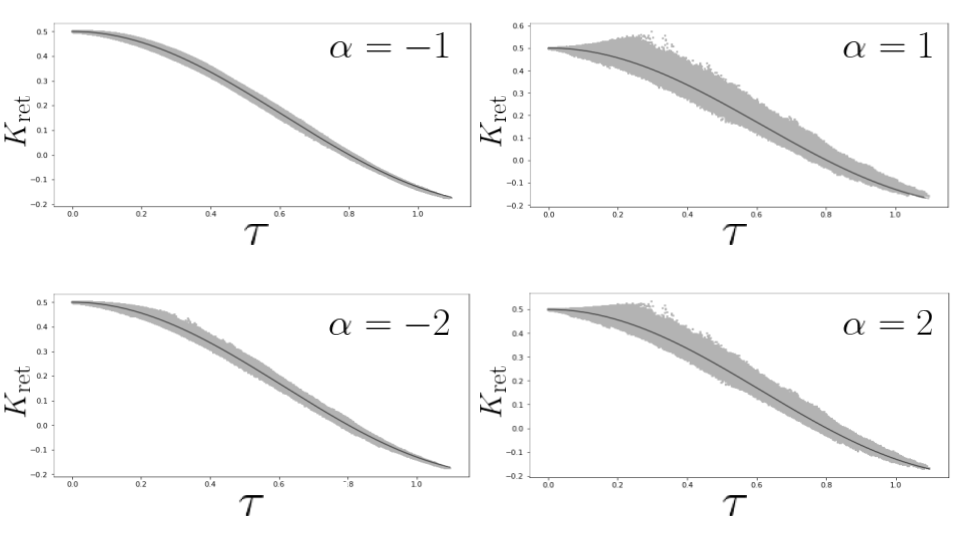}
    \caption{Numerical results for the causal set retarded propagator (grey) and the continuum retarded propagator (black) from a single sprinkling with $\rho = 4500$, $m = 3$, and various values of $\alpha$. The propagators are plotted as a function of the proper time measured in the manifold. The jump amplitudes in this calculation are from equation \ref{T Model} with the proper time estimated from the causal structure using equations \ref{proper time} and \ref{Js eqn}.}
    \label{fig:figure 4}
\end{figure*}

In figure \ref{fig:figure 4}, we see that this form of causal set propagator appears to agree with the continuum propagator at various values of $\alpha$. The variation also seems to decrease at larger proper times, though higher density simulations over larger causal diamonds may be necessary to state this conclusively.

\begin{figure}[ht]
    \centering
    \includegraphics[width=8.6cm]{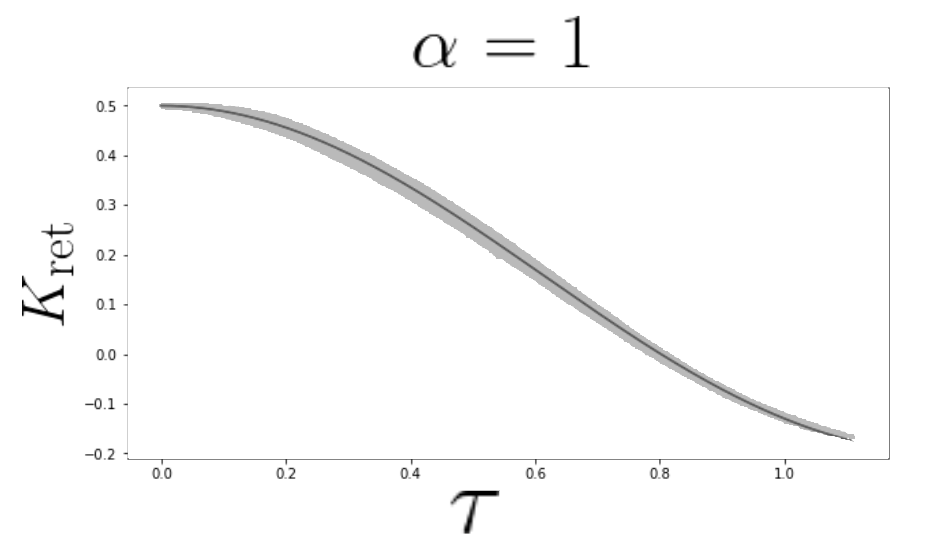}
    \caption{Numerical results for the causal set retarded propagator (grey) and the continuum retarded propagator (black) from a single sprinkling with $\rho = 4500$, $m = 3$, and $\alpha = 1$. The propagators are plotted as a function of the proper time measured in the manifold. The jump amplitudes in this calculation are from equation \ref{T Model} with the proper time calculated from the manifold.}
    \label{fig:figure 5}
\end{figure}

While the variation does seem to be larger for positive values of $\alpha$, there is good reason to believe this variation can be attributed to how we have estimated the proper times used in the jump amplitudes. Figure \ref{fig:figure 5} shows the same graph for the propagator with $\alpha = 1$, but in this case the jump amplitudes were calculated with the manifold proper time instead of the causal set estimate. As we can clearly see, the variation is significantly smaller. Note that in equation \ref{T Model}, positive values of $\alpha$ will make the jump amplitudes more sensitive to the proper time, so this variation is more noticeable.

Though we have only discussed integer values of $\alpha$ so far, the only limitation from the math is that $\alpha \neq 0$. Some conceptual problems do arise from allowing $\alpha << 1$. If $\alpha$ was sufficiently small that could cause $|T_{ij}| > 1$ for some $i,j$. While, in theory, the path sum should still average to the correct propagator, this situation would contradict the probability amplitude interpretation of $T$.

\begin{figure}[ht]
    \centering
    \includegraphics[width=8.6cm]{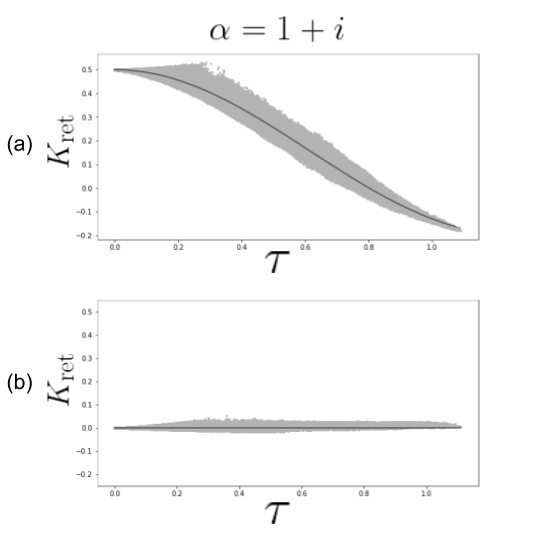}
    \caption{Numerical results for the causal set retarded propagator (grey) and the continuum retarded propagator (black) from a single sprinkling with $\rho = 4500$, $m = 3$, and $\alpha = 1 + i$. The propagators are plotted as a function of the proper time measured in the manifold. The jump amplitudes in this calculation are from equation \ref{T Model} with the proper time estimated from the causal structure using equations \ref{proper time} and \ref{Js eqn}. Plot (a) shows the real part of the propagator and plot (b) shows the imaginary part.}
    \label{fig:figure 6}
\end{figure}

We must also consider complex values of $\alpha$. $\alpha$ should be allowed to be complex because both the jump amplitudes $T$ and the propagator $K$ can be complex. Figure \ref{fig:figure 6} shows the results for the 2D retarded propagator using equation 38 with $\alpha = 1 + i$. As we can see, the causal set propagator is still a close match for the continuum value.

\subsection{\label{sec:results2} Results for Feynman Propagator}
The same process carried out to model the retarded propagator in the previous subsection can also be applied to create a path sum for the Feynman propagator. While past work such as~\cite{24} considered how one could obtain the Feynman propagator for a free scalar field from the retarded and advanced propagators, that work did not express the Feynman propagator as a path sum. This is useful to consider since, in a continuum calculation, the Feynman propagator can be directly calculated as a path integral but the retarded propagator cannot~\cite{10}.

In two dimensions~\cite{18}, the Feynman propagator is
\begin{eqnarray}
    K(y-x) = \Theta(\tau_{xy}^2)\left(-\frac{i}{4}H_0^{(2)}(m\tau_{xy})\right) \nonumber \\
    + \Theta(-\tau_{xy}^2)\left(\frac{1}{2\pi}K_0(ms_{xy})\right)
\end{eqnarray}
where $s_{xy} = \sqrt{-\tau_{xy}^2}$. Then, using equation \ref{T from K}, we must have the jump amplitude function
\begin{eqnarray}
    T(y-x) = \Theta(\tau_{xy}^2)\left(-\frac{i}{4a}H_0^{(2)}(\pm \beta m\tau_{xy})\right) \nonumber \\
    + \Theta(-\tau_{xy}^2)\left(\frac{1}{2\pi a}K_0(\pm \beta ms_{xy})\right)
\end{eqnarray}
Here, $H$ is a Hankel function and $K$ is a modified Bessel function. Note that since $K_0$ decays for large arguments, spacelike jumps far from the lightcone are very unlikely. By contrast, $H_0^{(2)}$ is oscillatory, so timelike jumps far from the lightcone should still be expected.

In four dimensions~\cite{18}, the Feynman propagator is
\begin{eqnarray}
    K(y-x) = \Theta(\tau_{xy}^2)\left(\frac{m}{8\pi \tau_{xy}}H_1^{(1)}(m\tau_{xy})\right) \nonumber \\
    + \Theta(-\tau_{xy}^2)\left(-\frac{im}{4\pi^2s_{xy}}K_1(ms_{xy})\right)
\end{eqnarray}
This yields the corresponding jump amplitude function
\begin{eqnarray}
    T(y-x) = \Theta(\tau_{xy}^2)\left(\frac{\pm \beta m}{8\pi a \tau_{xy}} H_1^{(1)}(\pm \beta m \tau_{xy})\right) \nonumber \\
    + \Theta(-\tau_{xy}^2) \left(-\frac{\pm i\beta m}{4\pi^2 a s_{xy}}K_1(\pm \beta m s_{xy})\right)
\end{eqnarray}

If we follow the same reasoning as with the retarded propagator, we can attempt to simplify the jump amplitudes in 4 dimensions by setting $a=-\rho/m^2$ which sets $\beta=0$. While this does not fully remove the $\tau$-dependence, it does greatly simplify the expression for $T(y-x)$. After taking the limit as $\beta \to 0$, we find
\begin{equation}
    T(y-x) = \frac{im^2}{4\pi^2\rho |\tau_{xy}^2|}
\end{equation}

One challenge is that this path sum would be non-local, which makes it difficult to verify these results numerically. A possible avenue for future work would be to simulate this numerically over large finite region of a flat spacetime to see if the results approach the continuum value for the propagator. Then additional work may be necessary to show that trajectories far from the lightcone do not contribute much. Alternatively, it may be necessary to include boundary terms that account for the paths far from the lightcone that cannot be modeled in a finite region.

\section{\label{sec:conclusion} Conclusion}
Constructing a propagator is a key step in establishing a field theory on a causal set. While past work has shown examples of path sums for propagators that match the continuum in the appropriate limits, these constructions were not unique solutions. In this paper, we have derived a general equation for how to relate the average value of a scalar field propagator to the possible average jump amplitudes. This enabled us to solve for the jump amplitudes necessary to recreate the correct continuum propagator on average. These were tested numerically in the case of the 2 dimensional retarded propagator and shown to match the continuum value. Even though these various constructions of the propagators should agree for large proper times, at small proper times they may differ greatly. This could make understanding possible values for jump amplitudes important for the small scale dynamics of a quantum field theory on a causal set.

\begin{acknowledgments}
I would like to acknowledge my advisor, Dr. David Craig for the many insightful conversations that shaped this paper. I would also like to acknowledge Dr. Steven Johnston for feedback about the content of the paper.
\end{acknowledgments}

\newpage

\bibliography{pathSums}

\end{document}